\documentstyle[aps,pra,epsf]{revtex}
\draft
\begin{document}

\title{Finite temperature theory of the scissors mode in a Bose
gas using the moment method}

\author{T. Nikuni}
\address{Department of Physics, University of Toronto, Toronto, Ontario,
Canada M5S 1A7} 

\date{\today}
\maketitle
\begin{abstract}
We use a generalized Gross-Pitaevskii equation for the condensate
and a semi-classical kinetic equation for the noncondensate atoms to
discuss the scissors mode in a trapped Bose-condensed gas
at finite temperatures.
Both equations include the effect of $C_{12}$ collisions
between the condensate and noncondensate atoms.
We solve the coupled moment equations describing oscillations
of the quadrupole moments of the condensate
and noncondensate components to
find the collective mode frequencies and collisional damping rates
as a function of temperature.
Our calculations extend those of Gu\'ery-Odelin and Stringari at
$T=0$ and in the normal phase.
They complement the numerical results of Jackson and Zaremba,
although Landau damping is left out of our approach.
Our results are also used to calculate the quadrupole response function,
which is related to the moment of inertia.
It is shown explicitly that the moment of inertia of a trapped Bose gas
at finite temperatures involves a sum of an irrotational component from the
condensate and a rotational component from the thermal cloud atoms.
\end{abstract}

\section{Introduction}
Superfluidity resulting from a Bose-Einstein condensate (BEC) in atomic gases is
characterized by the irrotational nature of the condensate flow.
The recent experimental observation of the scissors mode in a trapped Bose-condensed
gas \cite{Foot} clearly demonstrated the irrotational nature of a trapped
superfluid Bose gas.
The scissors mode is the oscillation of the atomic cloud with respect
to a symmetry axis of an anisotropic trap potential~\cite{GO},
induced by a sudden rotation of the trap about this axis.
Above the Bose-Einstein transition temperature ($T_{\rm BEC}$), the
thermal cloud exhibits two normal mode frequencies corresponding
to the rotational and irrotational motion.
In contrast, the pure condensate motion at $T=0$ only exhibits one frequency,
since only irrotational motion is allowed.
The above distinction between the condensate (superfluid) oscillation
at $T\sim 0$ and the thermal gas (normal fluid) oscillation
above $T_{\rm BEC}$ was clearly observed in the experiment reported in
Ref.~\cite{Foot}.
The observed frequencies of oscillations are in good agreement with the
theoretical predictions in Ref.~\cite{GO} at $T=0$ and $T>T_{\rm BEC}$.

At finite temperatures in the Bose-condensed phase, where an appreciable
fraction of the atoms are excited out of the condensate,
one expects coupled motions of the superfluid
and normal fluid components.
Recent experiments at Oxford \cite{Foot2} observed 
such a coupled scissors mode oscillations at finite temperatures,
and determined the temperature dependence of the frequency and damping rate
of the oscillations of each component.
In this paper, we discuss the scissors mode in a trapped Bose-condensed gas at
finite temperatures using the kinetic theory derived by
Zaremba, Nikuni and Griffin (ZNG) \cite{ZNG}.
In the ZNG kinetic theory, one has a generalized 
Gross-Pitaevskii equation for the condensate atoms and a
semi-classical kinetic equation for the noncondensate atoms.
The condensate and noncondensate are coupled through  mean-field interactions
as well as collisions between the atoms (the so-called $C_{12}$ collisions).
In this paper, we restrict ourselves to the collisionless regime,
where the mean collision rate
is much smaller than collective mode frequencies.
Generalizing the moment calculation approach used by Gu\'ery-Odelin and Stringari \cite{GO},
we derive coupled equations
describing oscillations of the quadrupole moments of the condensate and noncondensate
components at finite temperatures.

Recently, Jackson and Zaremba \cite{JZ} have solved the same coupled ZNG
equations numerically using FFT/Monte-Carlo simulations to discuss the
temperature-dependent oscillations associated with the scissors mode.
These authors found excellent agreement with the Oxford data \cite{Foot2}.
Our work is more analytical and complements the numerical results in Ref.\cite{JZ}.

\section{General moment method formalism}
We consider a Bose-condensed gas confined in an anisotropic harmonic trap potential
described by
\begin{equation}
U_{\rm ext}({\bf r})=\frac{m}{2}(\omega_x^2x^2+\omega_y^2y^2+\omega_z^2x^2),
\label{eq:trap}
\end{equation}
with $\omega_x^2=(1+\epsilon)\omega_0^2$ and $\omega_y^2=(1-\epsilon)\omega_0^2$.
The parameter $\epsilon$ characterizes the deformation of the trap potential in the
$x$-$y$ plane.
The coupled dynamics of the condensate and noncondensate~\cite{ZNG} is
described by the generalized Gross-Pitaevskii (GP) equation for the
condensate wavefunction $\Phi({\bf r},t)$
\begin{equation}
i\hbar\frac{\partial\Phi({\bf r},t)}{\partial t}=\left[-\frac{\hbar^2\nabla^2}{2m}
+U_{\rm ext}({\bf r})+gn_c({\bf r},t)+2g\tilde n({\bf r},t)-iR({\bf r},t)\right]
\Phi({\bf r},t), 
\label{eq:GP}
\end{equation}
and the semi-classical kinetic equation for the noncondensate distribution
function $f({\bf r},{\bf p},t)$
\begin{equation}
\frac{\partial f({\bf r},{\bf p},t)}{\partial t}+\frac{{\bf p}}{m}\cdot\bbox{\nabla}_{\bf r} 
f({\bf r},{\bf p},t)-\bbox{\nabla}U({\bf r},t)\cdot\bbox{\nabla}_{\bf p} f({\bf r},{\bf p},t)
=C_{12}[f,\Phi]+C_{22}[f].
\label{eq:QK}
\end{equation}
Here $n_c({\bf r},t)=|\Phi({\bf r},t)|^2$ is the condensate density,
and $\tilde n({\bf r},t)$ is the noncondensate density,
\begin{equation}
\tilde n({\bf r},t)=\int \frac{d{\bf p}}{(2\pi\hbar)^3} f({\bf r},{\bf p},t),
\label{eq:ntilde}
\end{equation}
and $U({\bf r},t)=U_{\rm ext}({\bf r})+2g[n_c({\bf r},t)+\tilde n({\bf r},t)]$
is the time-dependent effective potential acting on the noncondensate,
including the Hartree-Fock (HF) mean field.
As usual, we treat the interaction in the $s$-wave scattering approximation
$g=4\pi\hbar^2a/m$.
The dissipative term $R({\bf r},t)$ in the generalized GP equation (\ref{eq:GP})
is due to the collisional exchange of atoms in the condensate and noncondensate,
which is related to the $C_{12}$ collision integral in (\ref{eq:QK}),
\begin{equation}
R({\bf r},t)=\frac{\hbar\Gamma_{12}({\bf r},t) } {2n_c({\bf r},t)},~~
\Gamma_{12}({\bf r},t)=\int\frac{d{\bf p}}{(2\pi\hbar)^3}C_{12}[f({\bf r},{\bf p},t),
\Phi({\bf r},t)].
\label{R_term}
\end{equation}

The explicit expressions for the two collision integrals in the kinetic equation
(\ref{eq:QK}) are given by \cite{ZNG}
\begin{eqnarray}
C_{22}[f]&=&\frac{2g^2}{(2\pi)^5\hbar^7}\int d{\bf p}_2 \int d{\bf p}_3
\int d{\bf p}_3 \int d{\bf p}_4 \cr
&&\times\delta({\bf p}+{\bf p}_2-{\bf p}_3-{\bf p}_4)
\delta(\tilde\varepsilon_{p_1}+\tilde\varepsilon_{p_2}-\tilde\varepsilon_{p_3}
-\tilde\varepsilon_{p_4})\cr
&&\times [(1+f)(1+f_2)f_3f_4-ff_2(1+f_3)(1+f_4)],
\label{eq:C22}
\end{eqnarray}
\begin{eqnarray}
C_{12}[f,\Phi]&=&\frac{2g^2n_c}{(2\pi)^2\hbar^4}\int d{\bf p}_1 \int d{\bf p}_2
\int d{\bf p}_3 \cr
&&\times \delta (m{\bf v}_c+{\bf p}_1-{\bf p}_2-{\bf p}_3)
\delta(\varepsilon_c+\tilde\varepsilon_{p_1}-\tilde\varepsilon_{p_2}
-\tilde\varepsilon_{p_3}) \cr
&&\times[\delta({\bf p}_1-{\bf p})-\delta({\bf p}_2-{\bf p})-\delta({\bf p}_3-{\bf p})]
\cr && \times
[(1+f_1)f_2f_3-f_1(1+f_2)(1+f_3)].
\label{eq:C12}
\end{eqnarray}
Here $\tilde\varepsilon_p({\bf r},t)$ and $\varepsilon_c({\bf r},t)$ represent
the local energy of the noncondensate and condensate atoms
\begin{equation}
\tilde\varepsilon_p({\bf r},t)=\frac{p^2}{2m}+U({\bf r},t), ~~
\varepsilon_c({\bf r},t)=\frac{mv_c({\bf r},t)^2}{2}+\mu_c({\bf r},t),
\end{equation}
where the condensate chemical potential $\mu_c({\bf r},t)$ is defined
in (\ref{eq:muc}).

It is convenient to rewrite the GP equation in (\ref{eq:GP}) 
in terms of the amplitude
and phase variables $\Phi=\sqrt{n_c}e^{i\theta}$. 
This leads to the quantum hydrodynamic equations for the condensate:
\begin{mathletters}
\begin{eqnarray}
&&\frac{\partial n_c}{\partial t}+\bbox{\nabla}
\cdot(n_c {\bf v}_c)=-\Gamma_{12}, \label{hydro_nc}\\
&&m\frac{\partial {\bf v}_c}{\partial t}=-\bbox{\nabla}
\left(\mu_c+\frac{mv_c^2}{2}\right), \label{hydro_vc}
\end{eqnarray}
\label{hydro_c}
\end{mathletters}
where the condensate velocity is ${\bf v}_c({\bf r},t)\equiv
\hbar\bbox{\nabla}\theta({\bf r,t})/m$
and the condensate chemical potential $\mu_c({\bf r},t)$ is defined by
\begin{equation}
\mu_c({\bf r},t)\equiv -\frac{\hbar^2}{2m}
\frac{\nabla^2\sqrt{n_c({\bf r},t)}}
{\sqrt{n_c({\bf r},t)}}
+U_{\rm ext}({\bf r})+gn_c({\bf r},t)+2g\tilde n({\bf r},t).
\label{eq:muc}
\end{equation}
Throughout this paper, we use the Thomas-Fermi (TF) approximation, which
neglects the quantum pressure term in the condensate chemical potential to give:
\begin{equation}
\mu_c({\bf r},t)=U_{\rm ext}({\bf r})+gn_c({\bf r},t)+2\tilde n({\bf r},t).
\label{eq:TF}
\end{equation}
Within this TF approximation, the equilibrium condensate density profile
is given by
\begin{equation}
n_{c0}({\bf r})=\frac{1}{g}[\mu_{c0}-U_{\rm ext}({\bf r})]-2\tilde n_0({\bf r}).
\label{eq:nc0}
\end{equation}

The equilibrium distribution $f_0({\bf r},{\bf p})$ describing the noncondensate
atoms is given by the Bose-Einstein distribution function
\begin{equation}
f_0({\bf r},{\bf p})=\frac{1}{e^{\beta\left[\frac{p^2}{2m}+U_0({\bf r})-\mu_{c0}\right]
}-1},
\label{eq:f0}
\end{equation}
where $\beta=1/k_{\rm B}T$ and $U_0({\bf r})=U_{\rm ext}({\bf r})+2g[\tilde n_0({\bf r})
+n_{c0}({\bf r})]$.
The equilibrium density of the noncondensate described by (\ref{eq:f0}) is
\begin{equation}
\tilde n_0({\bf r})=\frac{1}{\Lambda^3}g_{3/2}(z_0({\bf r})),
\label{ntilde0}
\end{equation}
where $\Lambda=(2\pi\hbar^2/mk_{\rm B}T)^{1/2}$ is the thermal de Broglie
wavelength,
$g_{3/2}(z_0)$ is the Bose-Einstein function,
and $z_0=e^{\beta[\mu_{c0}-U_0({\bf r})]}$ is the local fugacity.
We note that the equilibrium TF densities of the condensate and noncondensate
determined from (\ref{eq:nc0}) - (\ref{ntilde0}) are isotropic in the
renormalized coordinates $(x',y',z')\equiv((\omega_x/\bar\omega) x,
(\omega_y/\bar\omega) y,(\omega_z/\bar\omega) z)$, where 
$\bar\omega\equiv (\omega_x\omega_y\omega_z)^{1/3}$.
Thus $n_{c0}({\bf r})$ and $\tilde n_0({\bf r})$ can be expressed as
functions of $r'=(x'^2+y'^2+z'^2)^{1/2}$.
Of course, the expression for $n_{c0}({\bf r})$ in (\ref{eq:nc0}) is
only valid within the TF condensate radius $r'<R'_{\rm TF}$, where
\begin{equation}
R'_{\rm TF}=\frac{2}{m {\bar\omega}^2}\left[\mu_{c0}-2g\frac{1}{\Lambda^3}
g_{3/2}(1)\right].
\end{equation}
Outside the TF condensate radius $r'>R'_{\rm TF}$, one has $n_{c0}=0$.

The scissors mode can be excited by a sudden rotation of the anisotropic trap
potential (\ref{eq:trap}) in the $x$-$y$ plane by a small angle $-\theta_0$.
This perturbation induces fluctuations in the quadrupole moment
\begin{equation}
Q(t)\equiv\int d{\bf r}~xy ~n({\bf r},t),
\end{equation}
where $n({\bf r},t)=n_c({\bf r},t)+\tilde n({\bf r},t)$.
Following Ref.~\cite{GO}, one can derive coupled moment equations 
involving the relevant quadrupole variables for the condensate and noncondensate,
starting from (\ref{hydro_c}) and (\ref{eq:QK}).
In the linearized theory, these exact moment equations are given by
(after some work)
\begin{mathletters}
\begin{eqnarray}
\frac{d}{dt}\langle xy \rangle_c&=&\langle xv_y+yv_x \rangle_c
-\langle xy \rangle_{12}, \label{xy_c}\\
\frac{d}{dt}\langle xv_y+yv_x\rangle_c&=&-2\omega_0^2 \langle xy \rangle_c
-\frac{2g}{m}\langle \bbox{\nabla}\tilde n_0 \cdot \bbox{\nabla}(xy) \rangle_c
+\frac{2g}{m}\langle \bbox{\nabla}n_{c0} \cdot \bbox{\nabla}(xy) \rangle_n, 
\label{xv_c}\\
\frac{d}{dt}\langle xy \rangle_n&=&\langle xv_y+yv_x \rangle_n
+\langle xy \rangle_{12}, \label{xy_n}\\
\frac{d}{dt}\langle xv_y+yv_x\rangle_n&=&-2\omega_0^2 \langle xy \rangle_n
-\frac{2g}{m}\langle \bbox{\nabla}n_{c0} \cdot \bbox{\nabla}(xy) \rangle_n
+\frac{2g}{m}\langle \bbox{\nabla}\tilde n_0 \cdot \bbox{\nabla}(xy) \rangle_c
+\frac{2}{m}\langle \bar P_{xy} \rangle_n, \label{xvp_n}\\
\frac{d}{dt}\langle xv_y-yv_x\rangle_n&=&-2\epsilon\omega_0^2 \langle xy \rangle_n
-\epsilon \frac{2g}{m}\langle \bbox{\nabla}n_{c0} \cdot \bbox{\nabla}(xy) \rangle_n
+\epsilon \frac{2g}{m}\langle \bbox{\nabla}\tilde n_0 \cdot \bbox{\nabla}(xy)
\rangle_c, \label{xvm_n}\\
\frac{d}{dt}\langle \bar P_{xy}\rangle_n&=&-m\omega_0^2
\langle xv_y+yv_x \rangle_n-m\epsilon\omega_0^2\langle xv_y-yv_x\rangle_n \cr
&&-\frac{2g}{m}\left\langle v_x\frac{\partial n_0}{\partial y}
+v_y\frac{\partial n_0}{\partial x} \right \rangle_n
+\left\langle \frac{p_xp_y}{m}\right\rangle_{\rm coll} \label{pxy_n}.
\end{eqnarray}
\label{moments_eq1}
\end{mathletters}
The various condensate and noncondensate moments related to some
function $\chi({\bf r})$ are defined by
\begin{equation}
\langle \chi \rangle_c\equiv \frac{1}{N}\int d{\bf r} n_c({\bf r},t)
\chi({\bf r}), ~~
\langle \chi \rangle_n\equiv \frac{1}{N}\int d{\bf r} \tilde n({\bf r},t)
\chi({\bf r}).
\end{equation}
\begin{equation}
\langle \chi{\bf v} \rangle_c\equiv \frac{1}{N}\int d{\bf r} 
n_c({\bf r},t)\chi({\bf r}){\bf v}_c({\bf r},t), ~~
\langle \chi {\bf v}\rangle_n\equiv \frac{1}{N}\int d{\bf r} 
\tilde n({\bf r},t)\chi({\bf r}){\bf v}_n({\bf r},t).
\end{equation}
Here the local velocity ${\bf v}_n$ and the local density of the
kinetic pressure tensor  $\bar P_{xy}$ of the thermal cloud are defined 
in terms of the single-particle distribution function by
\begin{equation}
\tilde n({\bf r},t){\bf v}_n\equiv\int \frac{d{\bf p}}{(2\pi\hbar)^3}
\frac{{\bf p}}{m} f({\bf r},{\bf p},t),
\end{equation}
\begin{equation}
\tilde n({\bf r},t)\bar P_{xy}({\bf r},t)\equiv\int \frac{d{\bf p}}{(2\pi\hbar)^3}
\frac{1}{m}(p_x-mv_{nx})(p_y-mv_{ny}) f({\bf r},{\bf p},t).
\end{equation}
In (\ref{moments_eq1}), one also has moments involving the collision integrals
$C_{12}$ and $C_{22}$:
\begin{equation}
\langle xy \rangle_{12}\equiv\frac{1}{N}\int d{\bf r}~xy \Gamma_{12}({\bf r},t),
\label{xy_12}
\end{equation}
\begin{equation}
\left\langle \frac{p_xp_y}{m} \right\rangle_{\rm coll}\equiv
\frac{1}{N}\int d{\bf r}\int \frac{d{\bf p}}{(2\pi\hbar)^3}
\frac{p_xp_y}{m}\left\{C_{12}[f({\bf r},{\bf p},t),\Phi({\bf r},t)]+
C_{22}[f({\bf r},{\bf p},t)]\right\}.
\label{pxy_coll}
\end{equation}
The condensate moments and noncondensate moments are 
coupled through the HF mean-field as well as the collisional exchange 
term $\langle xy \rangle_{12}$.
If we neglect the coupling between the two components
by setting $g=0$ and $\langle xy \rangle_{12}=0$ in (\ref{xy_c}) - (\ref{pxy_n}),
we recover the uncoupled moment equations derived in Ref.~\cite{GO},
namely the $T=0$ condensate equations in (\ref{xy_c}) and (\ref{xv_c}),
and the $T>T_{\rm BEC}$ thermal cloud gas equations in (\ref{xy_n}) - (\ref{pxy_n}).

\section{Truncated set of moment equations}

The moment equations in (\ref{moments_eq1}) are exact consequences of the
coupled ZNG equations (\ref{eq:GP}) and (\ref{eq:QK}), but
are obviously not closed because of the HF mean-field terms and the collision terms.
To obtain a closed set of equations, we must introduce an
ansatz for the condensate and noncondensate oscillations.
For the condensate, we assume that the velocity field has the
same irrotational form as the $T=0$ result, namely
\begin{equation}
{\bf v}_c({\bf r},t)=\alpha_c(t)\bbox{\nabla}(xy),
\label{ansatz_vc}
\end{equation}
where the small parameter $\alpha_c(t)$ characterizes the amplitude
of the condensate oscillation.
For the noncondensate atoms, it is convenient to write the
distribution function as
\begin{equation}
f({\bf r},{\bf p},t)=f_0({\bf r},{\bf p})+f_0({\bf r},{\bf p})
[1+f_0({\bf r},{\bf p})]\psi({\bf r},{\bf p},t),
\label{deltaf}
\end{equation}
where $\psi$ represents the small fluctuation of the noncondensate
around static equilibrium.
We make the ansatz that $\psi$ can be written in the form
\begin{equation}
\psi=\frac{\delta \tilde n({\bf r},t)}{\tilde \gamma_0({\bf r})}
-\beta{\bf v}_n({\bf r},t)\cdot{\bf p}+\alpha_{xy}(t)\beta \frac{p_xp_y}{m},
\label{ansatz_n}
\end{equation}
with $\tilde\gamma_0\equiv(\beta g/\Lambda^3)g_{1/2}(z_0)$ and $\beta=1/k_{\rm B}T$.
The first term in (\ref{ansatz_n}) is associated with the density fluctuations,
while the second term is associated with the velocity fluctuations.
We also assume that the noncondensate velocity field has the form
\begin{equation}
{\bf v}_n=\alpha_x y \hat {\bf x}+\alpha_y x \hat {\bf y}.
\label{ansatz_vn}
\end{equation}
The third term in (\ref{ansatz_n}) represents the quadrupole deformation
in the momentum distribution, characterized by a small parameter
$\alpha_{xy}(t)$.
The form of (\ref{ansatz_n}), with (\ref{ansatz_vn}),
is motivated by the exact solution \cite{KPS} of the
collisionless kinetic equation above $T_{\rm BEC}$ 
(Eq.~(\ref{eq:QK}) without $C_{12}$, $C_{22}$, and $2gn$),
These normal modes have frequencies $\omega_{\pm}=|\omega_x\pm\omega_y|$.

Taking the time derivative of (\ref{xv_c}), (\ref{xvp_n})
and (\ref{xvm_n}), and using (\ref{ansatz_vc}) and (\ref{ansatz_vn}),
we obtain the following set of equations:
\begin{mathletters}
\begin{eqnarray}
\frac{d^2}{dt^2}\langle xv_y+yv_x\rangle_c&=&
-\omega_0^2(2-\Delta_1)\langle xv_y+yv_x\rangle_c-\omega_0^2\Delta_2
\langle xv_y+yv_x \rangle_n \cr
&&+\frac{g}{m}\langle \bbox{\nabla}n_{c0}
\cdot\bbox{\nabla}(xy)\rangle_{12},  \label{moment_1}\\
\frac{d^2}{dt^2}\langle xv_y+yv_x\rangle_n&=&
-\omega_0^2(4+2\Delta_3-\Delta_2)\langle xv_y+yv_x\rangle_n
-\omega_0^2\Delta_1\langle xv_y+yv_x\rangle_c \cr
&&-2\epsilon\omega_0^2(1+\Delta_3)\langle xv_y-yv_x\rangle_n
-\frac{g}{m}
\langle \bbox{\nabla}n_{c0}\cdot\bbox{\nabla}(xy)\rangle_{12} \cr
&&+\frac{2}{m}\left\langle\frac{p_xp_y}{m}\right\rangle_{\rm coll}, 
\label{moment_2}\\
\frac{d^2}{dt^2}\langle xv_y-yv_x\rangle_n&=&
\epsilon\omega_0^2(2-\Delta_2)\langle xv_y+yv_x\rangle_n
+\epsilon\omega_0^2\Delta_1\langle xv_y+yv_x\rangle_c \cr
&&-\epsilon\frac{g}{m}\langle \bbox{\nabla}n_{c0}\cdot\bbox{\nabla}(xy)\rangle_{12}.
\label{moment_3}
\end{eqnarray}
\label{moments_eq2}
\end{mathletters}
The effect of the HF mean field is parameterized in (\ref{moment_1})-(\ref{moment_3})
by the following three quantities:
\begin{mathletters}
\begin{eqnarray}
\Delta_1&\equiv& \left\{\frac{2g}{\omega_0^2}\int d{\bf r}
[\bbox{\nabla}\tilde n_0\cdot\bbox{\nabla}(xy)]
[\bbox{\nabla}n_{c0}\cdot\bbox{\nabla}(xy)] \right\}/\Theta_{c,{\rm rigid}},
\label{delta1}
\\
\Delta_2&\equiv& \left\{\frac{2g}{\omega_0^2}\int d{\bf r}
[\bbox{\nabla}\tilde n_0\cdot\bbox{\nabla}(xy)]
[\bbox{\nabla}n_{c0}\cdot\bbox{\nabla}(xy)] \right\}/\Theta_{n,{\rm rigid}},
\label{delta2}
\\
\Delta_3&\equiv&\left[\frac{1}{1-\epsilon^2}\frac{2g}{\omega_0^2}
\int d{\bf r}\tilde n_0({\bf r}\cdot\bbox{\nabla}n_0)\right]
/\Theta_{n,{\rm rigid}} \cr
&=&\left[\frac{2}{(1-\epsilon^2)\tilde\omega_0^2}\int d{\bf r}\tilde P_0({\bf r})
\right]/\Theta_{n,{\rm rigid}}-1.
\label{delta3}
\end{eqnarray}
\end{mathletters}
Here $\tilde P_0\equiv (k_{\rm B}T/\Lambda^3)g_{5/2}(z_0({\bf r}))$ is the equilibrium
kinetic pressure, while
$\Theta_{c,{\rm rigid}}$ and $\Theta_{n,{\rm rigid}}$ are the rigid
values for the moment of inertia of the condensate and noncondensate
components (see, for example, Refs.\cite{ZS} and \cite{Stringari}):
\begin{eqnarray}
\Theta_{c,{\rm rigid}}&\equiv&m\int d{\bf r}(x^2+y^2)n_{c0}, \cr
\Theta_{n,{\rm rigid}}&\equiv&m\int d{\bf r}(x^2+y^2)\tilde n_0.
\label{Theta_nc}
\end{eqnarray}
For later use, we also note that (\ref{pxy_n}) can also be written as
\begin{equation}
\frac{d}{dt}\langle \bar P_{xy}\rangle_n
=-m\omega_0^2(1+\Delta_3)\left[\langle xv_y+yv_x\rangle_n
+\epsilon \langle xv_y-yv_x\rangle_n \right]
+\left\langle \frac{p_xp_y}{m}\right \rangle_{\rm coll}.
\label{pxy_n2}
\end{equation}
If we neglect the collisions, (\ref{moment_1})-(\ref{moment_3}) provide
a closed set of equations for the three moments $\langle xv_y+yv_x\rangle_c$,
$\langle xv_y+yv_x\rangle_n$, and $\langle xv_y-yv_x\rangle_n$,
which can be used to study the undamped scissors-mode oscillations
involving the condensate and noncondensate components at finite temperatures.

To truncate the collision terms $\langle xy \rangle_{12}$ and 
$\langle p_xp_y/m \rangle_{\rm coll}$ in (\ref{moments_eq2}),
we use (\ref{deltaf}) in the collision integrals and linearize in $\psi$
as follows [see also Ref.~\cite{LK}]:
\begin{eqnarray}
C_{22}[f]&\simeq&\frac{2g^2}{(2\pi)^5\hbar^7}\int d{\bf p}_2 \int d{\bf p}_3
\int d{\bf p}_4 \cr
&&\times \delta({\bf p}+{\bf p}_2-{\bf p}_3-{\bf p}_4) 
\delta(\tilde\varepsilon_p+\tilde\varepsilon_{p_2}-\tilde\varepsilon_{p_3}
+\tilde\varepsilon_{p_4}) \cr
&& \times f_0f_{20}(1+f_{30})(1+f_{40})(\psi_3+\psi_4-\psi_2-\psi)\equiv
\hat L_{22}[\psi],
\end{eqnarray}
and
\begin{eqnarray}
C_{12}[f,\Phi]&\simeq&
 \frac{2 g^2 n_c}{(2\pi)^2\hbar^4} \int d{\bf p}_1
\int d{\bf p}_2 \int d{\bf p}_3 \cr
&&\times  \delta(m{\bf v}_c+{\bf p}_1-{\bf p}_2-{\bf p}_3)
\delta(\mu_{c0}+\tilde\varepsilon_{p_1}
-\tilde\varepsilon_{p_2}-\tilde\varepsilon_{p_3}) \cr
&&\times  [\delta({\bf p}-{\bf p}_1)-\delta({\bf p}-{\bf p}_2)
-\delta({\bf p}-{\bf p}_3)] \cr
&&\times (1+f_{10})f_{20}f_{30}
(\beta g \delta n_c+\psi_2+\psi_3-\psi_1) \cr 
&& \equiv \beta g \delta n_c \hat L_{12}[1]+\hat L_{12}[\psi].
\label{C12_1}
\end{eqnarray}
Here the linearized $\hat L_{12}$ operator in (\ref{C12_1}) is defined by
\begin{eqnarray}
\hat L_{12}[\psi]
&\equiv&
 \frac{2 g^2 n_c}{(2\pi)^2\hbar^4} \int d{\bf p}_1
\int d{\bf p}_2 \int d{\bf p}_3 \cr
&&\times  \delta(m{\bf v}_c+{\bf p}_1-{\bf p}_2-{\bf p}_3)
\delta(\mu_{c0}+\tilde\varepsilon_{p_1}
-\tilde\varepsilon_{p_2}-\tilde\varepsilon_{p_3}) \cr
&&\times  [\delta({\bf p}-{\bf p}_1)-\delta({\bf p}-{\bf p}_2)
-\delta({\bf p}-{\bf p}_3)] \cr
&&\times (1+f_{10})f_{20}f_{30}
(\psi_2+\psi_3-\psi_1).
\label{L12}
\end{eqnarray}

Using the ansatz for $\psi$ given by (\ref{ansatz_n}), one finds that the
source function $\Gamma_{12}({\bf r},t)$ in (\ref{R_term}) reduces to
\begin{equation}
\Gamma_{12}({\bf r},t)=-\frac{\beta gn_{c0}({\bf r})}{\tau_{12}({\bf r})}
\left[\delta n_c({\bf r},t)+\frac{\delta \tilde n({\bf r},t)}{\tilde\gamma_0({\bf r})}
\right],
\label{G12_approx}
\end{equation}
where collision time $\tau_{12}$ is defined by
\begin{equation}
\frac{n_{c0}({\bf r})}{\tau_{12}({\bf r})}\equiv -\int \frac{d{\bf p}}{(2\pi\hbar)^3}
\hat L_{12}[1].
\end{equation}
Using (\ref{G12_approx}) and working to first order in $1/\tau_{12}$
(since we are working in the collisionless regime), one obtains the following equation 
of motion:
\begin{equation}
\frac{d}{dt}\langle \bbox{\nabla}n_{c0}\cdot\bbox{\nabla}(xy) \rangle_{12}=
\frac{m\omega_0^2}{g}\left[\frac{1}{\tau_1}\langle xv_y+yv_x\rangle_c
-\frac{1}{\tau_2}\langle xv_y+yv_x\rangle_n\right].
\label{dxy12dt}
\end{equation}
Here $\tau_1$ and $\tau_2$ are the averaged relaxation times associated
with the $C_{12}$ collisional exchange of atoms:
\begin{mathletters}
\begin{eqnarray}
\frac{1}{\tau_{1}}&=&\left\{
\frac{g}{\omega_0^2}\int d{\bf r}\frac{\beta gn_{c0}({\bf r})}
{\tau_{12}({\bf r})}[\bbox{\nabla}n_{c0}\cdot\bbox{\nabla}(xy)]^2\right\}
/\Theta_{c,{\rm rigid}},  \label{tau1}\\
\frac{1}{\tau_{2}}&=&\left\{
\frac{g}{\omega_0^2}\int d{\bf r}\frac{\beta gn_{c0}({\bf r})}
{\tau_{12}({\bf r})}[\bbox{\nabla}n_{c0}\cdot\bbox{\nabla}(xy)]^2\right\}
/\Theta_{n,{\rm rigid}}.
\label{tau2}
\end{eqnarray}
\end{mathletters}

One can also show that, using the ansatz for $f({\bf r},{\bf p},t)$ in (\ref{deltaf})
with (\ref{ansatz_n}), one can write $\langle p_xp_y/m\rangle_{\rm coll}$ defined by
(\ref{pxy_coll}) in the form of a relaxation time approximation
\begin{equation}
\left\langle \frac{p_xp_y}{m}\right\rangle_{\rm coll}
=-\frac{\langle\bar P_{xy}\rangle}{\tau_3},
\label{pxy_tau}
\end{equation}
with the associated relaxation time $\tau_3$ defined by
\begin{equation}
\frac{1}{\tau_3}\equiv \frac{\displaystyle \int d{\bf r} \frac{\tilde P_0({\bf r})}
{\tau_{\eta}({\bf r})}} {\displaystyle\int d{\bf r}\tilde P_0({\bf r})},
\end{equation}
where $\tau_{\eta}({\bf r})$ is the
viscous relaxation time (see also Refs.\cite{LK,relax}) defined by
\begin{equation}
\frac{1}{\tau_{\eta}({\bf r})}
=\frac{\displaystyle \int \frac{d{\bf p}}{(2\pi\hbar)^3}
p_xp_y\left\{\hat L_{12}[p_xp_y]+\hat L_{22}[p_xp_y]\right\}}
{\displaystyle \int\frac{d{\bf p}}{(2\pi\hbar)^3} (p_xp_y)^2f_0(1+f_0)}.
\end{equation}
Using (\ref{pxy_tau}) in (\ref{pxy_n2}), we find the following equation
of motion
$\langle p_xp_y/m\rangle_{\rm coll}$
\begin{equation}
\frac{d}{dt}\left\langle \frac{p_xp_y}{m}\right\rangle_{\rm coll}
=\frac{1}{\tau_3}m\omega_0^2(1+\Delta_3)
[\langle xv_y+yv_x\rangle_n+\epsilon \langle xv_y-yv_x\rangle_n]
-\frac{1}{\tau_3}\left\langle\frac{p_xp_y}{m}\right\rangle_{\rm coll}.
\label{pxydt}
\end{equation}

In summary, we have obtained a closed set of the three moment equations,
as given by (\ref{moments_eq2}), (\ref{dxy12dt}) and (\ref{pxydt}).
They involve three parameters $\Delta_i$ characterizing the HF
mean-field coupling and three collisional relaxation times
$\tau_i$.
One problem in actual calculations of these quantities is that
the integrands in the formulas for $\Delta_1$, $\Delta_2$,
$\tau_1$ and $\tau_2$ have singularities at the TF
condensate boundary (when $r'=R'_{\rm TF}$) arising from the divergence
in $\bbox{\nabla}\tilde n_0$ (we recall $g_{1/2}(1)\to\infty$).
These singular contributions are an unphysical artifact of the
simple TF approximation.
For our present purposes, we can
introduce a cutoff in the integrals where the
TF approximation breaks down.
More explicitly, we follow Refs.~\cite{DPS,LPS} and 
introduce a cutoff at $r'=R'_{\rm cutoff}$, where
\begin{equation}
R'_{\rm cutoff}=R'_{\rm TF}\left[1-\left(\frac{a_{\rm HO}}{2R_{\rm TF}}\right)^{4/3}
\right],
\end{equation}
with $a_{\rm HO}\equiv (\hbar/m\bar\omega)^{1/2}$.

It is convenient to introduce a 3-component vector describing the
dynamical moment variables:
\begin{equation}
{\bf w}=\left(\matrix{\langle xv_y+yv_x\rangle_c& \cr
\langle xv_y+yv_x\rangle_n& \cr
\langle xv_y-yv_x\rangle_n& \cr}\right),
\end{equation}
The coupled moment equations we have derived then reduce to the following matrix
equation
\begin{equation}
\frac{d^2}{dt^2}{\bf w}+
\omega_0^2\hat K_0 {\bf w}-\bbox{\gamma}=0.
\label{eq_matrix}
\end{equation}
The $3\times 3$ matrix $\hat K_0$ is given by
\begin{equation}
\hat K_0=\left ( \matrix{ 2-\Delta_1 & \Delta_2 &0 \cr
\Delta_1&4+2\Delta_3-\Delta_2 & 2\epsilon(1+\Delta_3)\cr
-\epsilon\Delta_1& -\epsilon(2-\Delta_2)& 0 \cr} 
\right ),
\end{equation}
and the column vector $\bbox{\gamma}$ describes the terms related
to collisions
\begin{equation}
\bbox{\gamma}=\left(\matrix{
\displaystyle{\frac{g}{m}}
\langle \bbox{\nabla}n_{c0}\cdot\bbox{\nabla}(xy)\rangle_{12}& \cr
-\displaystyle{\frac{g}{m}}
\langle \bbox{\nabla}n_{c0}\cdot\bbox{\nabla}(xy)\rangle_{12}
+{\displaystyle\frac{2}{m}}\left\langle\frac{p_xp_y}{m}
\right\rangle_{\rm coll}
& \cr
-\epsilon\displaystyle{\frac{g}{m}}
\langle \bbox{\nabla}n_{c0}\cdot\bbox{\nabla}(xy)\rangle_{12}
& \cr}\right).
\end{equation}
To first order in $1/\tau_i$, one has the following equation for
$\bbox{\gamma}$:
\begin{equation}
\frac{d}{dt}\bbox{\gamma}=\omega_0^3\hat K'{\bf w},
\end{equation}
where the elements of the $3\times 3$ matrix $\hat K'$ are given by
\begin{equation}
\hat K'=\left( \matrix{\displaystyle \frac{1}{\tilde\tau_1} &\displaystyle
 -\frac{1}{\tilde\tau_2} &0 \cr
\displaystyle-\frac{1}{\tilde\tau_1} & \displaystyle \frac{2}{\tilde\tau_3}+\frac{1}{\tilde\tau_2}
&\displaystyle \epsilon\frac{2}{\tilde\tau_3} \cr
-\displaystyle \epsilon\frac{1}{\tilde\tau_1} &
\displaystyle \epsilon\frac{1}{\tilde\tau_2} &0  \cr } 
\right).
\label{K'}
\end{equation}
In (\ref{K'}), we have introduced the renormalized relaxation
times
\begin{equation}
\frac{1}{\tilde\tau_1}\equiv \frac{1}{\omega_0\tau_1},~
\frac{1}{\tilde\tau_2}\equiv \frac{1}{\omega_0\tau_2},~
\frac{1}{\tilde\tau_3}\equiv \frac{1+\Delta_3}{\omega_0\tau_3}.
\end{equation}
We can now look for normal mode solutions of (\ref{eq_matrix})
with the time dependence
$e^{-i\omega t}$.
The matrix equation in (\ref{eq_matrix}) then reduces to
\begin{equation}
\left(\hat K_0-i\frac{1}{\tilde\omega}\hat K'\right){\bf w}=\tilde\omega^2{\bf w},
\label{eq_matrix2}
\end{equation}
where $\tilde \omega\equiv\omega/\omega_0$ is the renormalized mode frequency.
We discuss the solutions of (\ref{eq_matrix2}) in the next two sections. 

\section{Dynamics without collisions}
\subsection{Undamped oscillations of the two components}
In this section, we first discuss the collective oscillations of the condensate
and noncondensate components neglecting the collisional damping. 
This means we set $\bbox{\gamma}=0$ in (\ref{eq_matrix}), or equivalently,
$K'=0$ in (\ref{eq_matrix2}).
In this approximation, our eigenvalue equation reduces to
\begin{equation}
\hat K_0{\bf w}=\tilde\omega^2{\bf w}.
\label{eq_for_omega}
\end{equation}
Neglecting the HF mean fields ($\Delta_i=0$), the solution of (\ref{eq_for_omega})
is given by the zeros of the determinant
\begin{equation}
\left| \matrix{2-\tilde\omega^2 &0&0 \cr
              0 & 4-\tilde \omega^2 &2\epsilon \cr
              0 & -2\epsilon &-\tilde\omega^2 \cr} \right|
              =(2-\tilde\omega^2)(\tilde\omega^4-4\tilde\omega^2+4\epsilon^2)=0.
\end{equation}
In this simple limit, the mode frequencies are \cite{GO} $\omega_c=\sqrt{2}\omega_0$
and $\omega_{\pm}=\sqrt{2}\omega_0(1\pm\sqrt{1-\epsilon^2})^{1/2}$.
The thermal gas modes thus have a high frequency and a low frequency component.
Keeping $\Delta_i$ finite, we next solve the eigenvalue problem (\ref{eq_for_omega})
numerically, using the trap parameters of the recent Oxford experiment 
\cite{Foot2}, namely, $\omega_x=\omega_z=126$Hz, $\omega_y=\sqrt{8}\omega_x$.
The total number of atoms is taken to be $N=5\times 10^4$.

In Fig.~1, we plot the normal mode frequencies as a function of the temperature.
We find three normal mode frequencies, denoted by $\omega_c,\omega_+$ and $\omega_-$.
In the low temperature limit, $\omega_c$ approaches the well-known $T=0$ condensate mode 
frequency $\omega_c=\sqrt{2}\omega_0$ \cite{GO}. 
At finite temperatures, the frequency is shifted due to the HF mean field
of the noncondensate atoms.
Expanding the solution of (\ref{eq_for_omega}) to first order in the
interaction parameters $\Delta_i$, one finds
\begin{equation}
\omega_c^2\simeq  \omega_0^2(2-\Delta_1).
\label{omega_c}
\end{equation}
This formula for $\omega_c$ is a good approximation to the direct
numerical solution of (\ref{eq_for_omega}).
This result can also be obtained directly by neglecting the 
noncondensate fluctuations in (\ref{moment_1}), i.e.,
$\langle xv_y+yv_x\rangle_n=0$.
Thus the positive frequency shift in the condensate mode ($\omega_c$) 
in Fig.~1 is mainly due to the mean field of the static thermal cloud, $2g\tilde n_0$.
As for the noncondensate ($\omega_{\pm}$) modes, the frequencies approach the pure thermal
cloud frequencies found above $T_{\rm BEC}$ \cite{GO},
$\omega_{\pm}=|\omega_x\pm\omega_y|=\sqrt{2}\omega_0
(1\pm\sqrt{1-\epsilon^2})^{1/2}$ as $T$ approaches $T_{\rm BEC}$.
At low temperatures, however, there are large frequency shifts in these noncondensate
modes because of the coupling to a large condensate.

The solutions of (\ref{eq_for_omega}) also give the relative amplitude of the
quadrupole oscillations of the condensate and noncondensate components.
It is more convenient to use the rotation angles $\theta_c$ and $\theta_n$ 
of the condensate and noncondensate clouds,
assuming that the two components rotate without distortion.
These rotation angles are related to the quadrupole moments through
\begin{equation}
\theta_c=\langle xy \rangle_c/\langle x^2-y^2\rangle_{c0},~
\theta_n=\langle xy \rangle_n/\langle x^2-y^2\rangle_{n0}.
\label{theta_xy}
\end{equation}
We note that using the TF density profiles for $n_{c0}({\bf r})$ and
$\tilde n_0({\bf r})$ in (\ref{Theta_nc}), one obtains 
\begin{equation}
Nm\langle x^2-y^2\rangle_{c0}=-\epsilon\Theta_{c,{\rm rigid}},~~
Nm\langle x^2-y^2\rangle_{n0}=-\epsilon\Theta_{n,{\rm rigid}}.
\label{Q0TF}
\end{equation}
In Fig.2(a)-(c), we plot the temperature dependence of rotation amplitudes 
in (\ref{theta_xy}) of the condensate
and noncondensate oscillations associated with the $\omega_c$ and $\omega_{\pm}$ modes.
We find that the $\omega_c$ and $\omega_-$ modes involve in-phase oscillations
of the two components, while the $\omega_+$ mode involves out-of-phase oscillations.
At $T>0.5T_{\rm BEC}$, the $\omega_c$ mode mostly involves the condensate oscillations.
For decreasing $T$, the amplitude of the noncondensate component increases, and at low
temperatures, the thermal cloud is almost moving together with the condensate
cloud. In contrast, the $\omega_{\pm}$ modes involve mainly noncondensate oscillations
at all temperatures.
All these results agree with those in Ref.~\cite{JZ}.

We next consider the experimental situation \cite{Foot,Foot2} where the scissors modes
are excited by the sudden rotation of the trap potential by the
angle $-\theta_0$ at $t=0$.
We assume that the atoms are initially in thermal equilibrium for $t<0$.
The corresponding initial conditions for the angles  $\theta_c(t)$ and $\theta_n(t)$ 
are found from the moment equations in (\ref{moments_eq1}),
\begin{eqnarray}
\theta_c(0)&=&\theta_0,~~\theta_c''(0)=-2\omega_0^2\theta_0,~~
\theta_c''''(0)=4\omega_0^2\theta_0,~~ \cr
\theta_c'(0)&=&\theta_c'''(0)=\theta_c'''''(0)=0, \cr
\theta_n(0)&=&\theta_0,~~\theta_n''(0)=-2\omega_0^2\theta_0,~~
\theta_n''''(0)=4\omega_0^2[2+\Delta_3-(1+\Delta_3)\epsilon^2]\theta_0,~~ \cr
\theta_n'(0)&=&\theta_n'''(0)=\theta_n'''''(0)=0.
\label{initial}
\end{eqnarray}
The solution for $\theta_c(t)$ and $\theta_n(t)$ for $t>0$ with
the initial conditions in (\ref{initial}) is given by 
\begin{eqnarray}
\theta_c(t)/\theta_0&=&A_c\cos\omega_c t + A_+\cos\omega_+t+A_-\cos\omega_-t, \cr
\theta_n(t)/\theta_0&=&B_n\cos\omega_n t + B_+\cos\omega_+t+B_-\cos\omega_-t,
\label{angles}
\end{eqnarray}
where the amplitudes of each mode are given by the following explicit formulas:
\begin{eqnarray}
A_i&=&
\frac{\omega_j^2\omega_k^2-2\omega_0^2(\omega_j^2+\omega_k^2)
+4\omega_0^4}{(\omega_i^2-\omega_j^2)(\omega_i^2-\omega_k^2)}, \cr
B_i&=&
\frac{\omega_j^2\omega_k^2-2\omega_0^2(\omega_j^2+\omega_k^2)
+4\omega_0^4[(2+\Delta_3)-(1+\Delta_3)\epsilon^2]}
{(\omega_i^2-\omega_j^2)(\omega_i^2-\omega_k^2)}.
\label{coeff}
\end{eqnarray}
In (\ref{coeff}), $i,j,k$ represent the three mode indices ($c,+,-$)
and $i\neq j\neq k$.
In Fig. 3, we plot the temperature dependence of these amplitudes $A_i$ and $B_i$.
We find that the oscillations of the condensate always have a large weight
in the condensate mode ($\omega_c$).
In contrast, the thermal cloud oscillations make a large contribution to
the noncondensate modes ($\omega_{\pm}$) at $T>0.5T_{\rm BEC}$,
but the condensate mode ($\omega_c$) is dominant at low temperatures.
This behavior is in agreement with the numerical results obtained by
Jackson and Zaremba \cite{JZ}.

\subsection{Quadrupole response function and the moment of inertia}
Our results for the scissors mode can be used to derive an expression for the moment of
inertia, a quantity which characterizes the rotational properties of a trapped Bose gas.
Zambelli and Stringari \cite{ZS} derived the following exact relation between the
moment of inertia $\Theta$ and the quadrupole response function:
\begin{equation}
\Theta=-\frac{4\epsilon^2m}{\pi}
\int d\omega \frac{\chi_Q''(\omega)}{\omega^3},
\label{Theta}
\end{equation}
where $\chi_Q''(\omega)$ is the imaginary part of the quadrupole
response function of the system.
One can also show that $\chi_Q''(\omega)$ is related to the Fourier transform
of the quadrupole moment induced by the sudden rotation of the
trap potential,
\begin{equation}
\chi''_Q(\omega)=-\frac{\pi}{\lambda}\omega Q(\omega),~~\lambda\equiv
-2\theta_0\epsilon m\omega_0^2.
\label{chi_Q}
\end{equation}
Using the solution for $\theta_c(t)$ and $\theta_n(t)$ in
(\ref{angles}) with the relations in (\ref{theta_xy}),
we find that $Q(\omega)$ is given by
\begin{eqnarray}
Q(\omega) 
&=&-\frac{\epsilon}{2m}\left\{
\Theta_{c,{\rm rigid}}
\sum_i A_i[\delta(\omega-\omega_i)+\delta(\omega+\omega_i)] 
\right. \cr
&&\left. +\Theta_{n,{\rm rigid}}
\sum_i B_i[\delta(\omega-\omega_i)+\delta(\omega+\omega_i)]
\right\}.
\label{Q_omega}
\end{eqnarray}
One can show that $\chi_Q''(\omega)$ given by (\ref{chi_Q}) with (\ref{Q_omega})
satisfies the exact sum rule relation \cite{ZS}
\begin{equation}
\int d\omega \chi''_Q(\omega)\omega=-(\pi/m)\Theta_{\rm rigid}.
\end{equation}
Here $\Theta_{\rm rigid}\equiv\Theta_{c,{\rm rigid}}+\Theta_{n,{\rm rigid}}$ is the rigid
value of the moment of inertia of the total system (see (\ref{Theta_nc})),
\begin{equation}
\Theta_{\rm rigid}=m\int d{\bf r}(x^2+y^2)n_0({\bf r}),
\label{Theta_rig}
\end{equation}
where $n_0({\bf r})=n_{c0}({\bf r})+\tilde n_0({\bf r})$.

Using these results for $\chi''_Q(\omega)$ in (\ref{Theta}), one finds
\begin{equation}
\Theta=2\epsilon^2\left(\Theta_{c,{\rm rigid}}\sum_i\frac{A_i}{\tilde\omega_i^2}
+\Theta_{n,{\rm rigid}}\sum_i\frac{B_i}{\tilde\omega_i^2}\right).
\label{m_inertia0}
\end{equation}
From the explicit formulas for $A_i$ and $B_i$ given in (\ref{coeff}), we find
\begin{equation}
2\epsilon^2\sum_i\frac{A_i}{\tilde\omega_i^2}=\epsilon^2
-\frac{(1-\epsilon^2)\Delta_1}
{(2-\Delta_1-\Delta_2)}, ~~
2\epsilon^2\sum_i\frac{B_i}{\tilde\omega_i^2}=1+\frac{(1-\epsilon^2)\Delta_2}
{(2-\Delta_1-\Delta_2)}.
\end{equation}
Noting that $\Theta_{c,{\rm rigid}}\Delta_1=\Theta_{n,{\rm rigid}}\Delta_2$
which follows from the definitions of $\Delta_1$ and $\Delta_2$ in (\ref{delta1})
and (\ref{delta2}),
we finally obtain the following simple formula for moment of inertia:
\begin{equation}
\Theta=\epsilon^2\Theta_{c,{\rm rigid}}+\Theta_{n,{\rm rigid}}.
\label{m_inertia}
\end{equation}
It is interesting to note that because of the interactions between the
two components, the irrotational contribution $\epsilon^2\Theta_{c,{\rm rigid}}$
and the rotational contribution $\Theta_{n,{\rm rigid}}$ in (\ref{m_inertia})
cannot be associated separately with the condensate and noncondensate quadrupole
moments in (\ref{m_inertia0}).
Thus obtaining the simple final expression given in (\ref{m_inertia}) is nontrivial.
In Fig.~4, we plot the moment of inertia in
(\ref{m_inertia}) as a function of temperature, as well as
the separate irrotational contribution $\epsilon^2\Theta_{c,{\rm rigid}}$
and the rotational contribution $\Theta_{n,{\rm rigid}}$.
An expression analogous to (\ref{m_inertia}) for the noninteracting trapped
Bose gas  was obtained by Stringari (see especially Fig.~1 of Ref.~\cite{Stringari}).

The above result for the moment of inertia in (\ref{m_inertia}) can be
understood by considering the anisotropic trap potential in (\ref{eq:trap})
rotating with the angular frequency $\Omega$ in the $x-y$ plane.
In this rotating frame, the stationary distribution of the noncondensate
is given by \cite{JZNG}
\begin{equation}
f_{\Omega}({\bf r},{\bf p})=\frac{1}{\exp\{\beta[p^2/2m+U({\bf r})-\Omega{\bf p}\cdot
(\hat{\bf z}\times{\bf r})-\tilde\mu]\}-1}.
\end{equation}
To first order in $\Omega$,
the total angular momentum of the noncondensate component is given by
[see also Ref.~\cite{GPS}]
\begin{eqnarray}
L_{n,z}&=&\int d{\bf r}\int\frac{d{\bf p}}{(2\pi\hbar)^3}(xp_y-yp_x)
f_{\Omega}({\bf r},{\bf p}) \cr
&\simeq&\Omega\int d{\bf r}
\int\frac{d{\bf p}}{(2\pi\hbar)^3}(xp_y-yp_x)^2f_0(1+f_0) \cr
&=&\Omega\int d{\bf r} (x^2+y^2)
\int\frac{d{\bf p}}{(2\pi\hbar)^3}\frac{p^2}{3}f_0(1+f_0) \cr
&=&\Omega  \int d{\bf r} ~m(x^2+y^2) \tilde n_0({\bf r}) \cr
&=&\Omega\Theta_{n,{\rm rigid}}.
\end{eqnarray}
Thus the noncondensate part of the moment of inertia is given by the
rigid value defined in (\ref{Theta_nc}), i.e., 
$\Theta_n\equiv L_{n,z}/\Omega=\Theta_{n,{\rm rigid}}$.

For the condensate contribution, one has to find the stationary solution of the GP
equation in the rotating frame \cite{ZS}.
In the linear regime, the stationary velocity field is determined from
\begin{equation}
\bbox{\nabla}[n_{c0}({\bf v}_c-{\bf \Omega}\times{\bf r})]=0.
\end{equation}
For the equilibrium TF condensate density $n_{c0}$ determined from 
(\ref{eq:nc0}), one can show that the stationary velocity field
is given by
\begin{equation}
{\bf v}_c=-\epsilon\Omega\bbox{\nabla}(xy).
\end{equation}
The total angular momentum of the condensate component is then given by
[see (\ref{Q0TF})]
\begin{eqnarray}
L_{c,z}&=&\int d{\bf r} ~m\left({\bf r}\times{\bf v}_c n_{c0}({\bf r})\right) \cr
&=&-\epsilon\Omega\int d{\bf r}~ m(x^2-y^2)n_{c0}({\bf r}) \cr
&=&\epsilon^2\Omega\Theta_{c,{\rm rigid}}.
\end{eqnarray}
Thus the condensate part of the moment of inertia is given by
$\Theta_c\equiv L_{c,z}/\Omega=\epsilon^2\Theta_{c,{\rm rigid}}$.
This has the same form as the $T=0$ expression for the irrotational
value of the moment of inertia of the TF condensate given in Ref.~\cite{ZS}. 
Thus, the first term in (\ref{m_inertia}) represents the irrotational
value of the condensate component in the Thomas-Fermi limit,
while the second term is the rigid value of the noncondensate component.
The expression in (\ref{m_inertia}) is analogous to the non-interacting
gas expression ($\Delta_i=0$) first obtained by Stringari \cite{Stringari}.
However, as we noted in the derivation of (\ref{m_inertia}), this
equivalence is nontrivial when $\Delta_i\neq 0$ because of the cancellation
of contributions in going from (\ref{m_inertia0}) to (\ref{m_inertia}).

\section{damping due to collisions}
We next consider the damping of the collective modes due to the collisions between atoms.
Since we are working in the collisionless regime $\omega\tau_i\gg 1$,
in (\ref{eq_matrix2}), we can treat $\hat K'$ as a perturbation to $\hat K_0$.
Working to first order in $1/\tau_i$, one has the following equation for
normal mode frequencies:
\begin{equation}
\det\left(\hat K_0-\frac{i}{\tilde\omega}\hat K'-\tilde\omega^2\hat 1\right)
\simeq
\det(\hat K_0-\tilde\omega^2\hat 1)-i\frac{1}{\tilde\omega}f_{\tau}(\tilde\omega)=0,
\label{det2}
\end{equation}
where $\hat 1$ is $3\times 3$ unit matrix and 
the function $f_{\tau}(\tilde\omega)$ related to damping is found to be given by
\begin{eqnarray}
f_{\tau}(\tilde\omega)&\equiv& \frac{1}{\tilde\tau_1}\left[
\tilde\omega^4-(4+2\Delta_3-\Delta_2)\tilde\omega^2+2\epsilon^2(2-\Delta_2)
(1+\Delta_3)\right] \cr
&&+(\tilde\omega^2-2+\Delta_1)\left\{\frac{2}{\tilde\tau_3}
[\tilde\omega^2-2\epsilon^2(1-\Delta_2/2)]+\frac{1}{\tilde\tau_2}
[\tilde\omega^2-2\epsilon^2(1+\Delta_3)]\right\} \cr
&&-2\epsilon^2\frac{\Delta_1\Delta_2}{\tilde\tau_3}
-2\left[\tilde\omega^2-2\epsilon^2(1+\Delta_3)\right]\frac{\Delta_2}{\tilde\tau_1}.
\end{eqnarray}
We note that the lowest order or first term in (\ref{det2}) can be written in terms
of the undamped frequencies $\tilde\omega_i$ determined from (\ref{eq_for_omega}),
namely
\begin{equation}
\det(\hat K_0-\tilde\omega^2\hat 1)=(\tilde\omega_c^2-\tilde\omega^2)
(\tilde\omega_+^2-\tilde\omega^2)(\tilde\omega_-^2-\tilde\omega^2).
\end{equation}
We now look for the damping of these frequencies, i.e.,
\begin{equation}
\tilde\omega=\tilde\omega_i-i\tilde\Gamma_i.
\end{equation}
To first order in $1/\tau_i$, we find the following expression for the damping rates:
\begin{equation}
\tilde\Gamma_i=\frac{f_{\tau}(\tilde\omega_i)}{2\tilde\omega_i^2(\tilde\omega_i^2
-\tilde\omega_j^2)(\tilde\omega_i^2-\tilde\omega_k^2)},~~~
(i\neq j\neq k).
\label{gamma}
\end{equation}

Expanding the solution (\ref{gamma}) in $\Delta_i$, one can obtain
approximate expressions for the damping rate $\Gamma_i$.
The lowest-order approximation (neglecting $\Delta_1,\Delta_2$ and $\Delta_3$)
gives after some algebra
\begin{equation}
\Gamma_c=\frac{1}{4\tau_1},~~\Gamma_{\pm}=\frac{1}{4\tau_3}+\frac{1}{8\tau_2}.
\label{gamma_approx}
\end{equation}
We recall that within this approximation, $\omega_c=\sqrt{2}\omega_0$ and 
$\omega_{\pm}=\sqrt{2}\omega_0(1\pm\sqrt{1-\epsilon^2})^{1/2}$.
One can show that damping for the condensate mode given in (\ref{gamma_approx})
is precisely equivalent to that obtained by Williams and Griffin \cite{WG},
where the dynamics of the noncondensate was neglected.
To see this, we recall the ansatz (\ref{ansatz_vc}) for ${\bf v}_c$ and
rewrite $\Gamma_c$ given in (\ref{gamma_approx}) as
[see the definition of $1/\tau_1$ in (\ref{tau1})]
\begin{equation}
\Gamma_c=\frac{1}{2\omega_c^2} \frac{\displaystyle \int d{\bf r}
\frac{\beta n_{c0}}{\tau_{12}} [g\bbox{\nabla}\cdot(n_{c0}{\bf v}_c)]^2 }
{\displaystyle m\int d{\bf r}~n_{c0} v_c^2}.
\label{gamma_c}
\end{equation}
Using the quantum hydrodynamic equations for the condensate in 
(\ref{hydro_c}) (neglecting $\delta\Gamma_{12}$ and $\delta\tilde n$),
one can write (\ref{gamma_c}) in terms of the density fluctuations
$\delta n_c$ in the undamped solution.
It is then straightforward to show that (\ref{gamma_c}) is equivalent
to the damping in Eq.~(21) of Ref.~\cite{WG}.

We have numerically evaluated (\ref{gamma}) to find the damping of the collective modes.
In Fig.~4, we plot the temperature dependence of the damping rates.
It turns out that the lowest-order formula for $\Gamma_c$ given
in (\ref{gamma_approx}) gives a good approximation
for the full expression in (\ref{gamma}).
This means that the collisional damping of the condensate mode
is almost entirely due to the $C_{12}$ collisional exchange with the static
thermal cloud, as first discussed in Ref.~\cite{WG},
with little contribution from the $C_{22}$ collision term in (\ref{pxy_coll}).
Since the viscous relaxation only involves noncondensate fluctuations,
they only affect the condensate motion through the HF mean field
$2g \tilde n$. 
Thus the contribution of $\tau_3$ to the condensate damping $\Gamma_c$
is of second order in $\Delta_i$, which is a small correction.

In contrast, the damping of the $\omega_{\pm}$ modes is mainly due to the viscous
relaxation associated with the $\tau_3$ relaxation time.
The increase of $\Gamma_{\pm}$ as the temperature decreases near $T_{\rm BEC}$
is due to the increase in the $C_{12}$ collision rate.
This clearly indicates that the equilibration of the thermal cloud in a
trapped Bose-condensed gas is enhanced by the $C_{12}$ collisions with
the condensate cloud \cite{relax}.
Above $T_{\rm BEC}$, one does not have $C_{12}$ collisions and thus the damping
$\Gamma_{\pm}$ is given by (\ref{gamma_approx}) without the
$1/\tau_2$ term, while the $\tau_3$ relaxation time now only involves
the contribution from the $C_{22}$ collisions \cite{GO}.

The damping of the condensate mode we have found is much
smaller than the magnitude observed in the Oxford experiments \cite{Foot2}.
This is because the present moment analysis does not include Landau damping 
of the condensate mode.
Landau damping is due to the mean-field coupling of the condensate
oscillations to a continuum of noncondensate excitations \cite{GP}.
In our moment calculation, we truncate the HF mean field coupling using a
simple ansatz for the noncondensate fluctuations, as given in (\ref{ansatz_n}).
This simple form describes the coherent motion of the thermal cloud,
but completely neglects the effect from incoherent single-particle excitations
which are responsible for Landau damping.
Recently, Jackson and Zaremba \cite{JZ} discussed the scissors mode at finite
temperatures using a direct Monte-Carlo simulation of the ZNG semi-classical
kinetic equation.
They found that for the same parameters used in the present study,
Landau damping is the main contribution to the damping of the condensate
mode at all temperatures (see also Ref.~\cite{WG}). 
Moreover, in Ref.~\cite{JZ}, the $C_{22}$ collisions appear to mainly influence 
the damping by relaxing the system towards equilibrium and hence allowing
the Landau mechanism to be more effective.
The $C_{12}$ contribution to the damping of the condensate mode in Ref.~\cite{JZ}
is essentially in agreement with our results for the $C_{12}$ collisional
damping rate in Fig.~5.
We note that the $C_{12}$ collisional damping of the condensate mode will 
increase in importance with the total number of atoms in the trapped gas.

We also note that the numerical results in Ref.~\cite{JZ} find an
increase of the damping of the noncondensate modes with decreasing
temperature.
This is in contrast to our result for the collisional damping rate
$\Gamma_{\pm}$ shown in Fig.~5.
The large damping in the noncondensate modes found in Ref.~\cite{JZ} at
low temperatures is presumably
due to the anharmonicity in the effective potential $U_0({\bf r})$
for the noncondensate, arising from the large condensate mean field
$2gn_{c0}({\bf r})$ \cite{Zaremba}.
This effect is not included in our moment calculation because of
our simple ansatz for the noncondensate fluctuations, given in (\ref{ansatz_n}).

\section{conclusions}
In this paper, we have used the ZNG kinetic theory \cite{ZNG}
to study the scissors mode
in a trapped Bose-condensed gas at finite temperatures.
We derived three coupled moment equations starting from a generalized
GP equation for the condensate and a semi-classical kinetic equation
for the noncondensate.
Our coupled moment equations are the natural extension 
to finite temperatures of a similar calculation carried out in Ref.~\cite{GO} 
for $T=0$ and $T>T_{\rm BEC}$.
We solved our coupled moment equations to find normal mode solutions.
These solutions were used to calculate the quadrupole response function
and the moment of inertia, using the exact frequency moment sum rule
formulas derived by Zambelli and Stringari~\cite{ZS}.
It was shown that the moment of inertia is given by a simple expression
in (\ref{m_inertia}), sum of the
irrotational value of the condensate component and the rotational
value of the noncondensate component.
This result clearly shows that the linear response of an
trapped Bose gas at finite temperatures involves the irrotational
motion of the condensate component as well as the rotational motion
of the noncondensate component.

We also evaluated the collisional damping for both the condensate and
noncondensate oscillations. 
Our analysis naturally led to three collision times:
(1)$\tau_1$ is the relaxation time of the condensate due to $C_{12}$
collisions.
(2)$\tau_2$ is the relaxation time of the noncondensate due to $C_{12}$ collisions.
(3)$\tau_3$ is the relaxation time of the noncondensate similar to
that involved in the shear viscosity transport coefficient \cite{relax},
which has a contribution from both $C_{12}$ and $C_{22}$ collisions.
We found that the noncondensate oscillations were damped mainly by a
viscous-type relaxation ($1/\tau_3$), while the condensate damping was dominated by
the collisional exchange due to $C_{12}$ collisions ($1/\tau_1$).
If we neglect the dynamical mean-field coupling between the condensate
and the noncondensate, one obtains an explicit expression for the condensate
damping, which is identical to that given in Ref.~\cite{WG} assuming a static noncondensate.
We find that this simple static thermal cloud approximation provides a quantitatively good
approximation for the $C_{12}$ collisional damping of the condensate mode.

However, our present discussion does not explain the damping observed
in the Oxford experiments \cite{Foot2}.
As discussed in Section V, this is because a truncated moment calculation
such as ours cannot easily deal with the mechanism of Landau damping, which is the dominant
contribution to the collective mode damping in the collisionless or mean-field
regime \cite{JZ,WG,GP}.
In order to incorporate Landau damping in the moment method,
one will have to solve the exact set of moment equations (\ref{xy_c}) - (\ref{pxy_n})
including the incoherent motion of the thermal cloud.
Our simple ansatz in (\ref{ansatz_n}) only describes the coherent thermal
cloud dynamics.

In spite of the limitations mentioned above, 
the moment method developed in this paper is very useful in understanding the
detailed dynamics of a trapped Bose-condensed gas at finite temperatures.
It provides an explicit physical picture of the
coupled motions of the condensate and noncondensate components involved
in the three scissors modes.
Our work was inspired by, and complements, the recent numerical work of
Jackson and Zaremba \cite{JZ} on the scissors mode.
Our more analytical theory is essentially consistent with
the results given in Ref.\cite{JZ}, apart from missing the Landau damping.
Our method can be easily extended to study other collective modes, such as
the monopole mode and the quadrupole modes \cite{Jamie}, as well as to study the
effect of a rotating thermal cloud at finite temperatures in vortex nucleation \cite{JZNG}.

\bigskip
\begin{center}
{\bf ACKNOWLEDGMENTS}
\end{center}
\bigskip
I thank A. Griffin for useful suggestions and a critical reading of the manuscript.
I also have benefited from stimulating discussions with E. Zaremba, 
B. Jackson and J.E. Williams.
This research was supported by JSPS of Japan.

\noindent
\clearpage

\begin{figure}
\epsfxsize=110mm
\centerline{\epsfbox{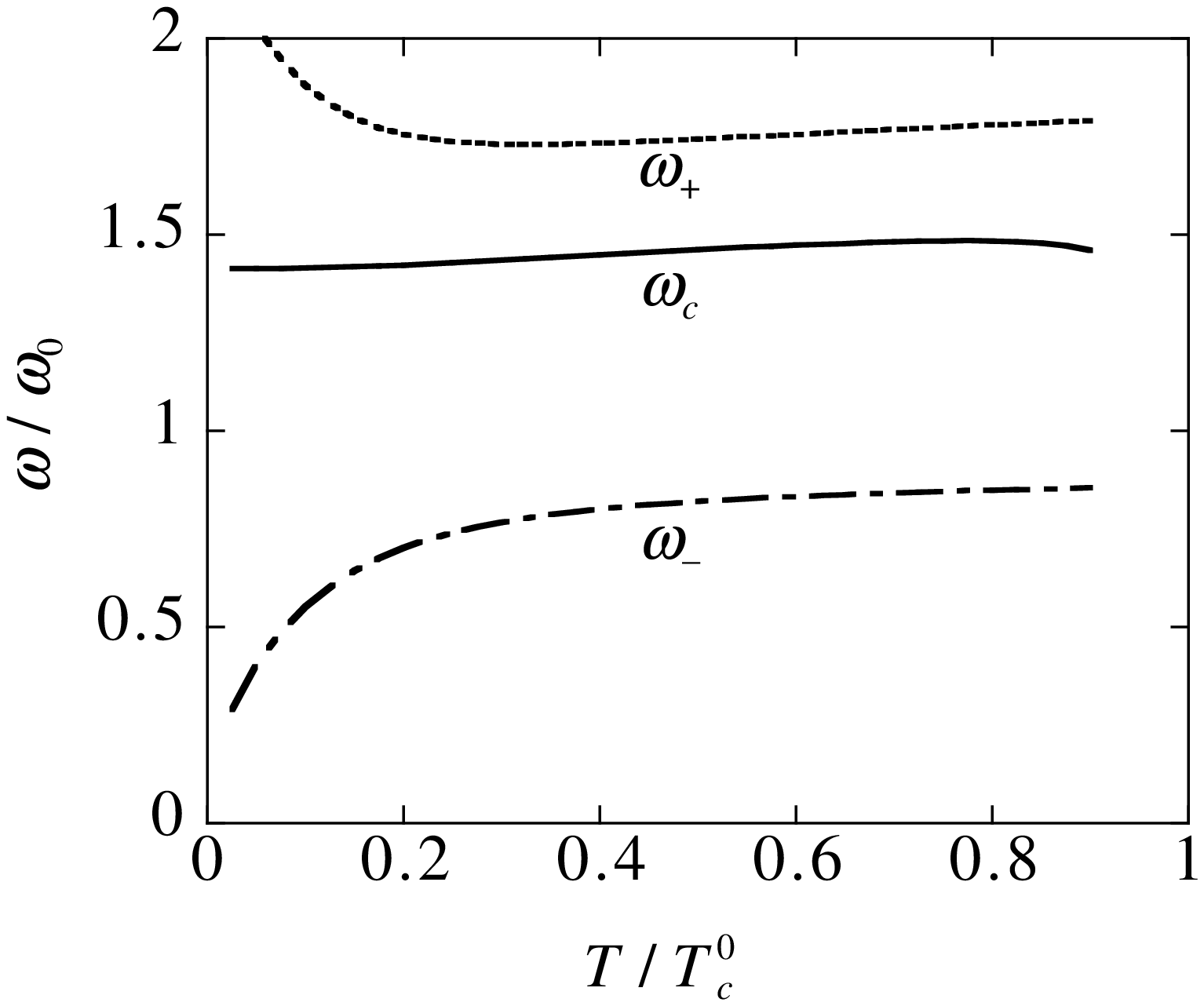}}
\begin{caption}
{Calculated temperature dependence of three scissors-mode frequencies for
$N=5\times 10^4$ $^{87}$Rb atoms in an anisotropic trap with
$\omega_x=\omega_z=126$Hz and $\omega_y=\sqrt{8}\omega_x$ 
{\protect\cite{Foot,Foot2}}.
The frequency is renormalized by $\omega_0=[(\omega_x^2+\omega_y^2)/2]^{1/2}$
and the temperature is renormalized by the ideal gas transition temperature
$T_c^0\equiv (\hbar\bar\omega/k_{\rm B})[N/g_3(1)]^{1/3}$.
The actual transition temperature is $T_{\rm BEC}\simeq 0.93 T_c^0$.}
\end{caption}
\label{fig1}
\end{figure}

\clearpage

\begin{figure}
\epsfxsize=110mm
\centerline{\epsfbox{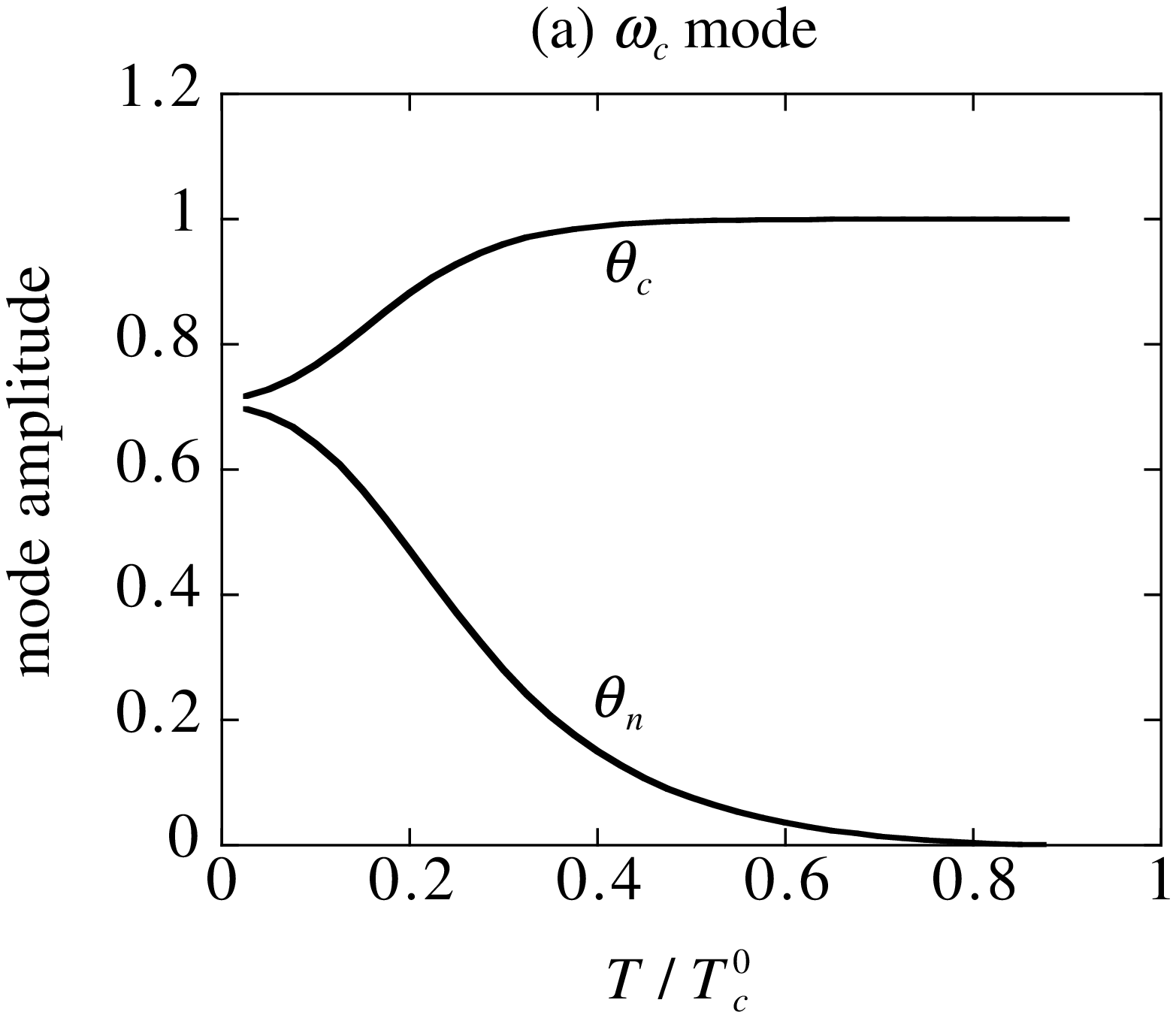}}
\epsfxsize=110mm
\centerline{\epsfbox{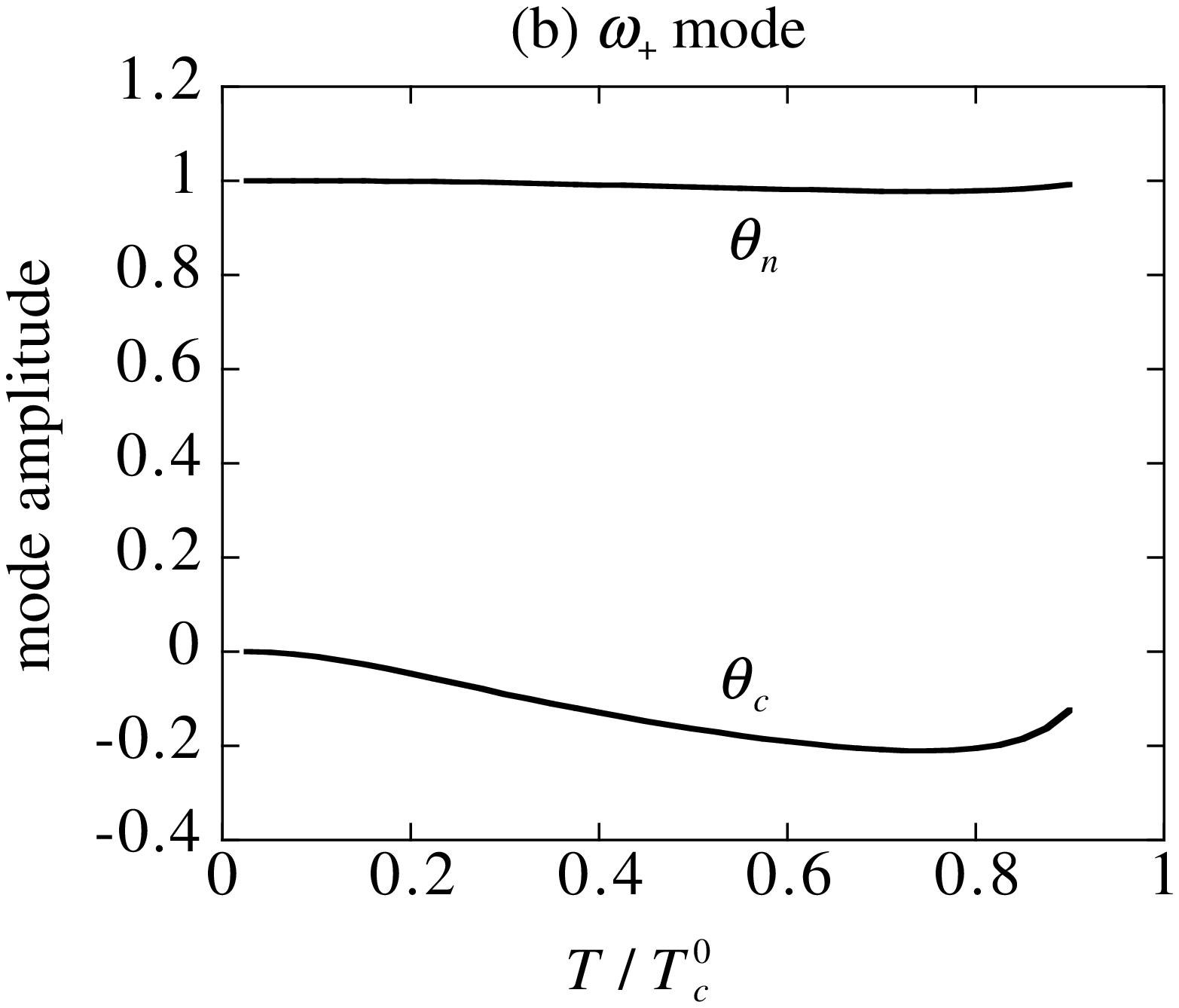}}
\epsfxsize=110mm
\centerline{\epsfbox{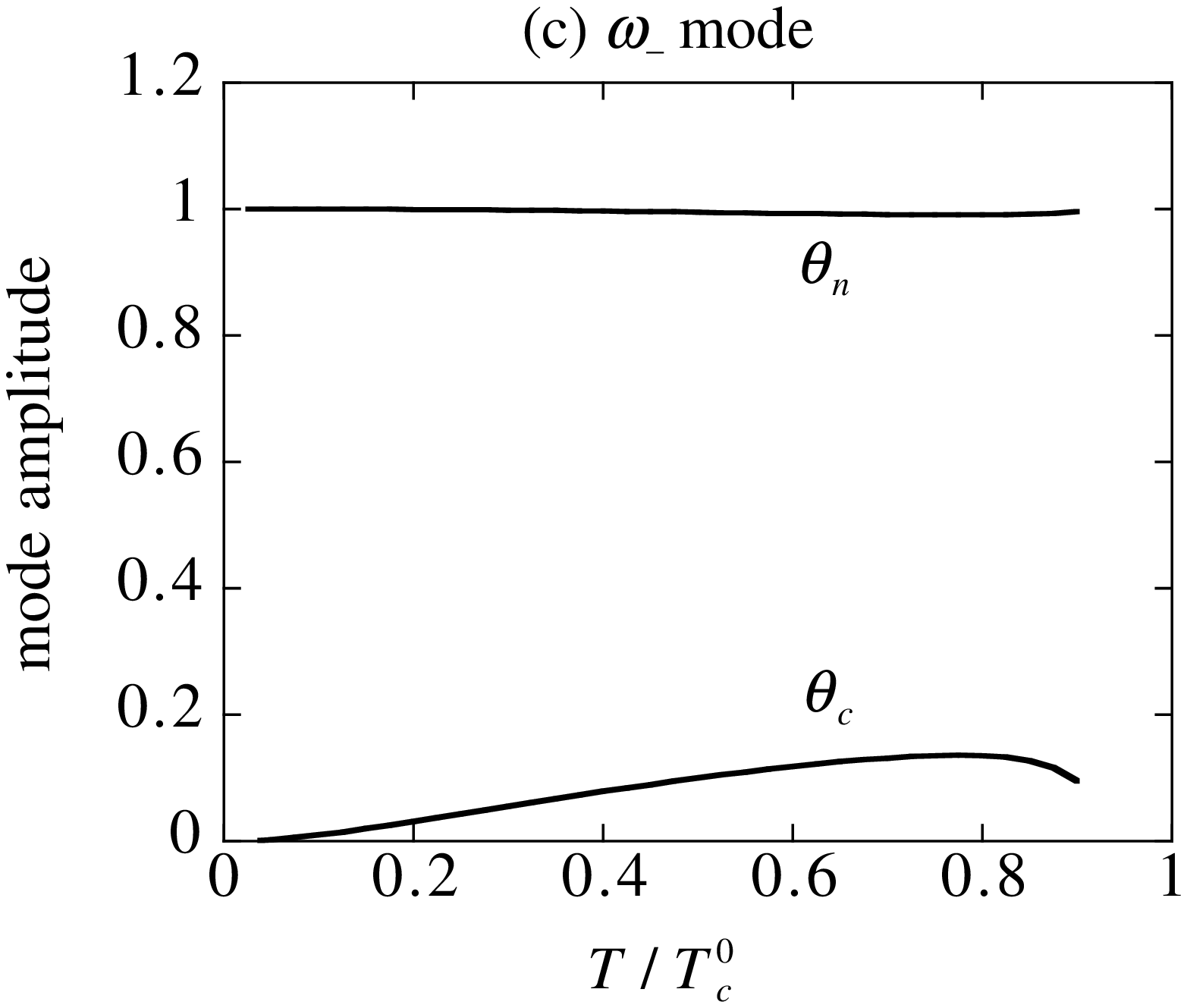}}
\begin{caption}
{Temperature dependence of the rotation angles $\theta$ of the two components
associated with (a) $\omega_c$ mode, (b) $\omega_+$ mode, 
and (c) $\omega_-$ mode.}
\end{caption}
\label{fig2}
\end{figure}

\clearpage

\begin{figure}
\epsfxsize=110mm
\centerline{\epsfbox{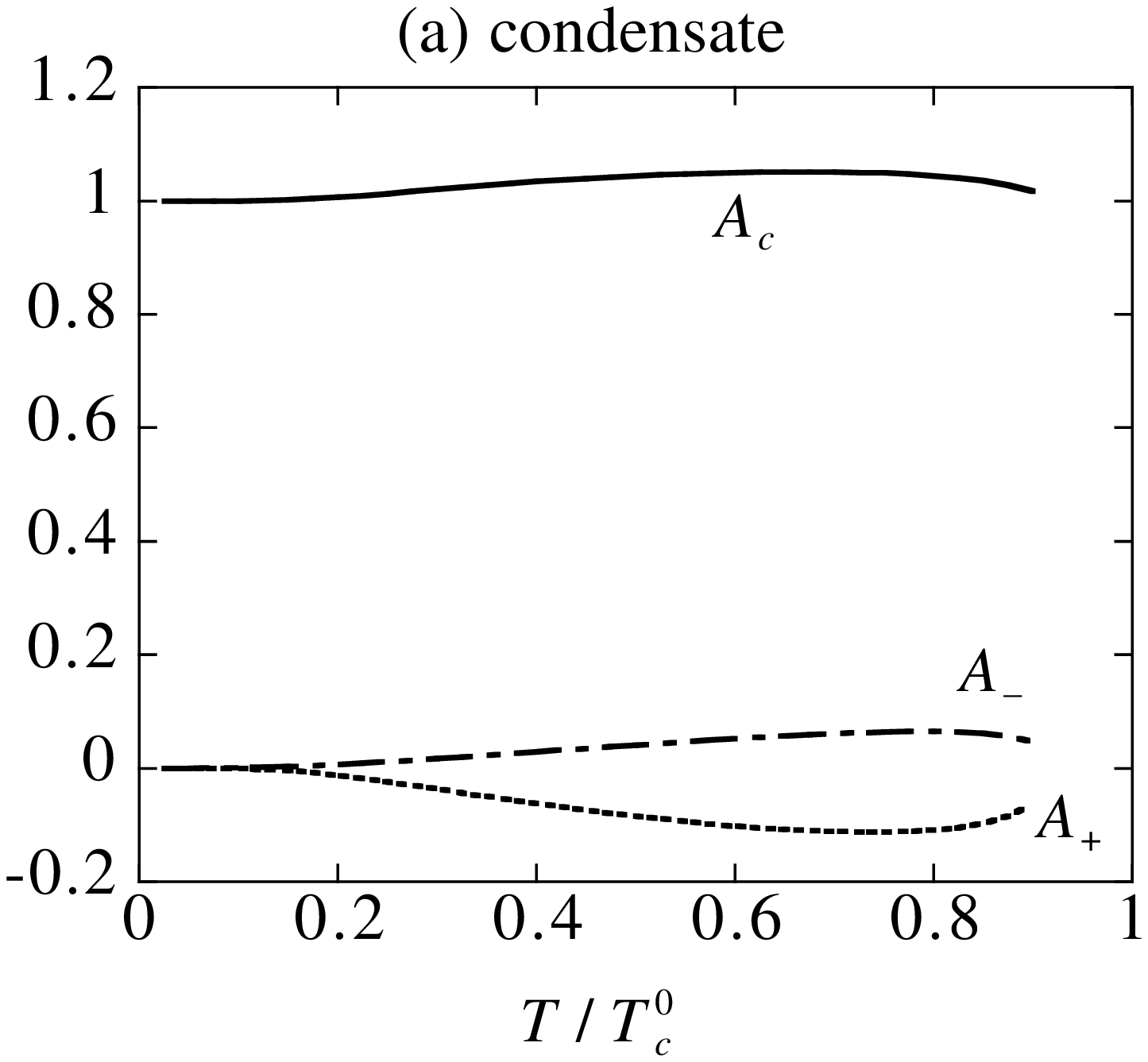}}
\epsfxsize=110mm
\centerline{\epsfbox{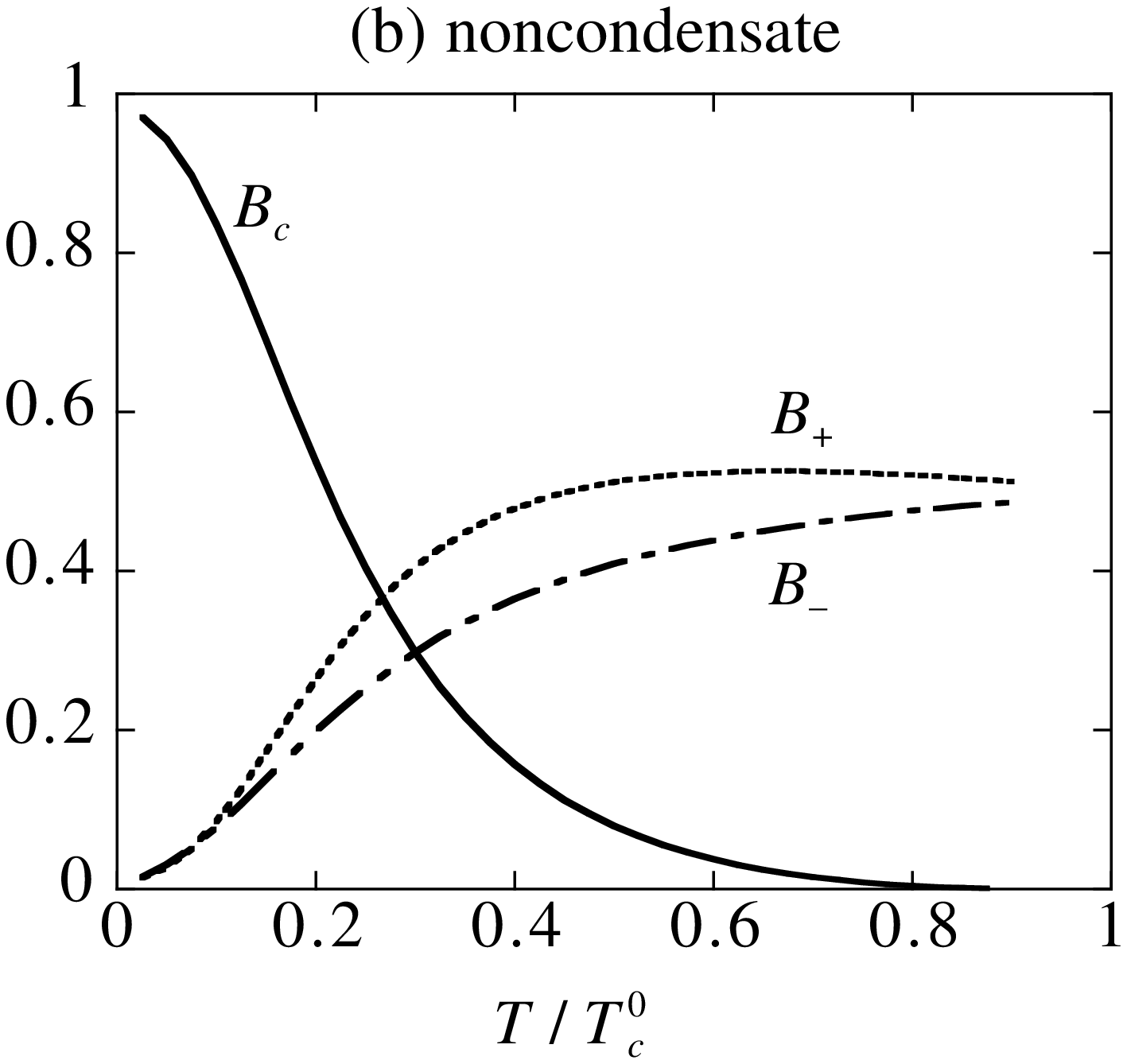}}
\begin{caption}
{Temperature dependence of coefficients $A_i$ and $B_i$ 
in (\ref{angles}) and (\ref{coeff}) for coupled scissors-mode 
oscillations induced by the sudden rotation of the trap potential.
Panel (a) is a plot of the condensate rotation angle ($A_i$) and panel (b) is a plot of the 
noncondensate angle ($B_i$), as defined in (\ref{angles}).}
\end{caption}
\label{fig3}
\end{figure}

\begin{figure}
\epsfxsize=110mm
\centerline{\epsfbox{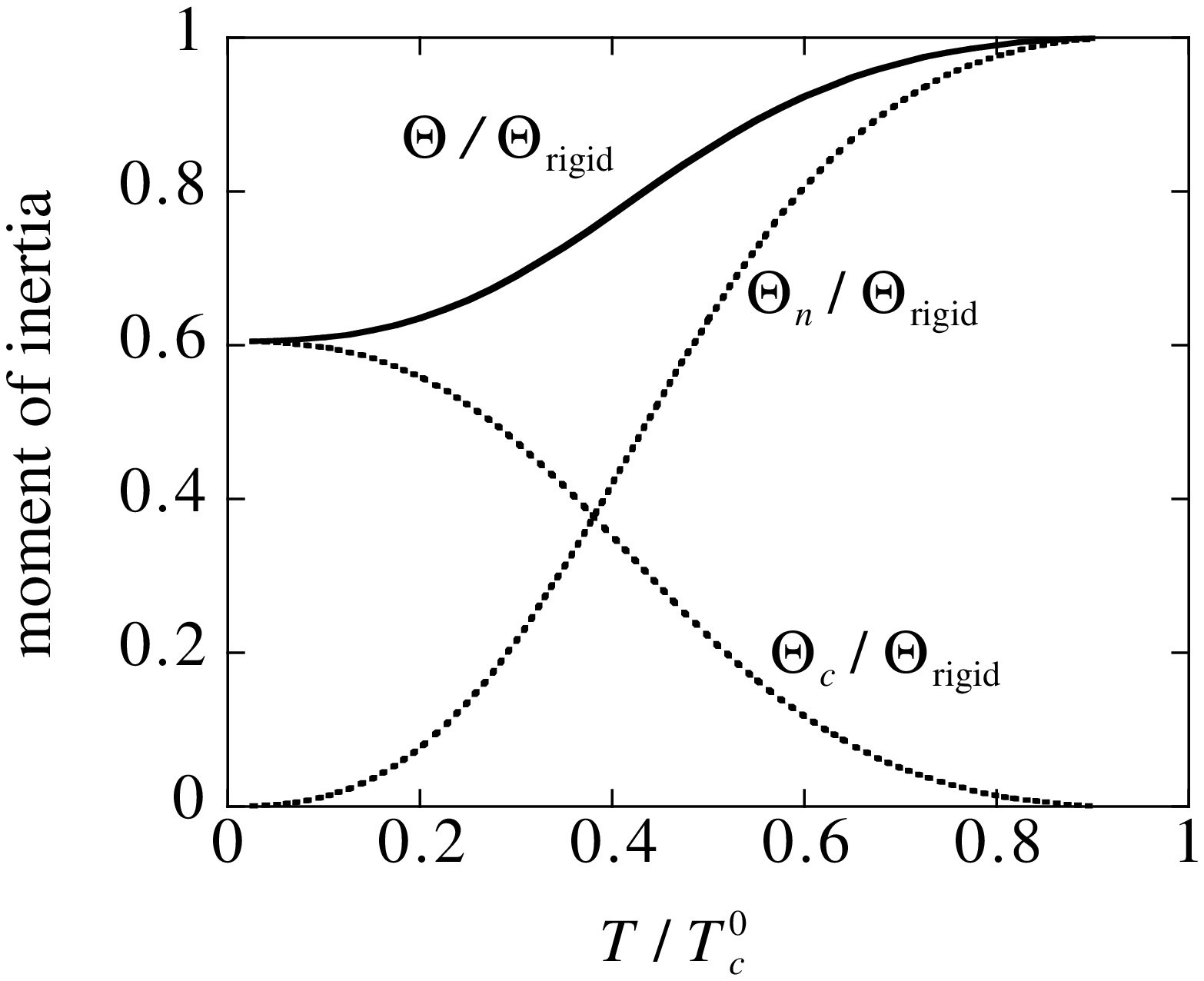}}
\begin{caption}
{Temperature dependence of the moment of inertia $\Theta$ in
(\ref{m_inertia}), normalized to $\Theta_{\rm rigid}$ defined by (\ref{Theta_rig}).
We also plot the irrotational condensate component
$\Theta_c=\epsilon^2\Theta_{c,{\rm rigid}}$
and the rotational noncondensate component $\Theta_n=\Theta_{n,{\rm rigid}}$,
where $\Theta_{c,{\rm rigid}}$ and $\Theta_{n,{\rm rigid}}$ are defined
in (\ref{Theta_nc}).}
\end{caption}
\end{figure}

\begin{figure}
\epsfxsize=110mm
\centerline{\epsfbox{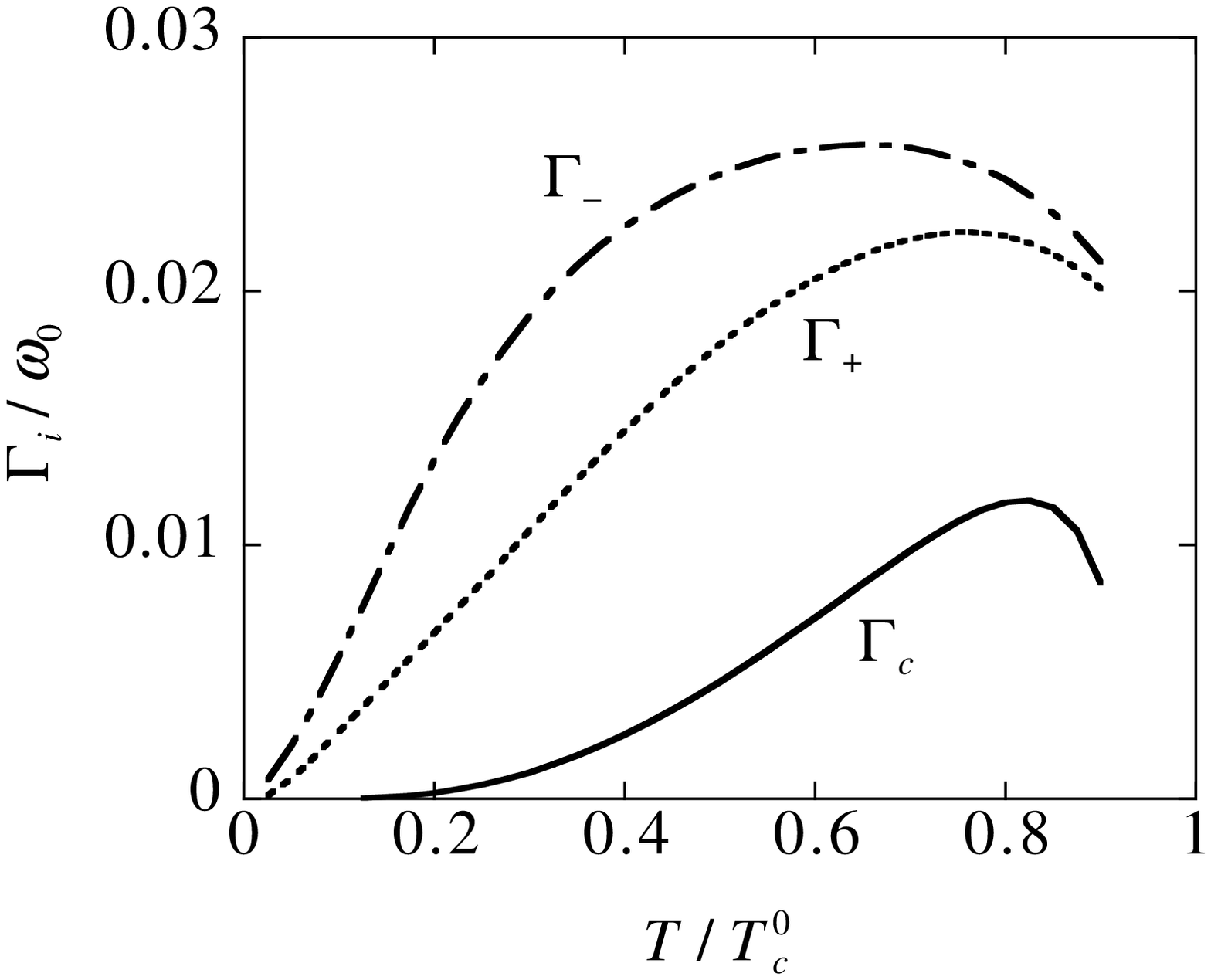}}
\begin{caption}
{Temperature dependence of the collisional damping rates associated with
three scissors-mode frequencies given by (\ref{gamma}).
As discussed in the text, Landau damping is not included.}
\end{caption}
\end{figure}

\end{document}